\documentclass[10pt,numberedappendix]{emulateapj}
\usepackage{booktabs}
\usepackage{graphicx}
\usepackage{dcolumn}
\usepackage{bm}
\usepackage{amsmath}

\usepackage{color}

\def\nar{\ref@jnl{New A Rev.}}          

\def\msun{$M_\odot$}

\begin{document}

\title{Finding the One: Identifying the Host Galaxies of
  Gravitational-Wave Sources}

\author{Hsin-Yu Chen$^1$ and Daniel E. Holz$^2$}
\affiliation{$^1$Department of Astronomy and Astrophysics, University of
  Chicago, Chicago, Illinois 60637, USA\\
$^2$Enrico Fermi Institute, Department of Physics and Kavli Institute
for Cosmological Physics\\University of Chicago, Chicago, Illinois 60637, USA}

\begin{abstract}
We explore the localization of compact binary coalescences with ground-based gravitational-wave detector networks. We
simulate tens of thousands of binary events, and present the distributions of
localization sky areas and localization volumes for 
a range of sources and network configurations. We show that generically there exists a tail
of particularly well-localized events, with 2D and 3D localizations of
$<10\,\mbox{deg}^2$ and $<1000\,\mbox{Mpc}^3$ achievable, respectively, starting
in LIGO/Virgo's third observing run. Incorporating estimates for the galaxy
density and the binary event rates, we argue that future gravitational-wave
detector networks will localize a small number of binary
systems per year to a sufficiently small volume that the unique host
galaxy might be identified. For these golden events, which are
generally the closest and loudest ones, the gravitational-wave detector
networks will point (in 3D; the length of the finger matters) directly at the
source. This will allow for studies of the properties of the host
galaxies of compact binary mergers, which may be an important component in
exploring the formation channels of these sources. In addition, since the host
will provide an independent measurement of the redshift, this will allow the use
of the event as a standard siren to measure cosmology.
Furthermore, identification of a small number of host galaxies can enable deep follow-up
searches for associated electromagnetic transients.
\end{abstract}

\maketitle

\section{Introduction}
\label{sec:intro}
The era of gravitational-wave astrophysics has
arrived~\citep{2016PhRvL.116f1102A,2016PhRvL.116x1103A}, and we now
await multi-messenger astronomy to achieve the full
scientific potential of these detections.
The observation of an electromagnetic\footnote{Under ``electromagnetic'' we also include
particle messengers, such as neutrinos and cosmic rays.} (EM)
counterpart to a gravitational-wave (GW) source would allow us to more fully characterize and
understand the physics and astrophysics of the sources.
In addition to direct EM counterparts to the GW events, there is also enormous
interest in identifying the host galaxies to these events. For example, 
measurements of the stellar age, mass, or metallicity of host galaxies, perhaps
even as a function of redshift or binary total mass, could lead to major
insights and constraints regarding astrophysical formation mechanisms~\citep{2016ApJ...818L..22A}.
Furthermore, host galaxies will provide independent estimates of redshift,
allowing for the use of gravitational-wave sources as standard sirens~\citep{1986Natur.323..310S}.
Finally, by identifying a host galaxy we dramatically increase the
probability of identifying a transient counterpart associated with the event,
since we are then able to utilize sensitive narrow-field instruments to search.

There are two generic ways to identify a host galaxy to a GW event. The first is by detecting
a transient counterpart that can be directly associated with the
gravitational-wave event. For example, a contemporaneous short gamma-ray burst in a
consistent area of the sky and at a consistent distance would be strong evidence
for a direct association between the gravitational-wave and the electromagnetic
sources. Once the transient is detected, one can often directly identify the
host galaxy. This
approach has generated enormous interest, spawning a very active EM follow-up
community~\citep{2016ApJ...826L..13A,2016ApJ...823L..33S,2016ApJ...823L..34A,2016ApJ...826L..29C}.

An alternate
way to identify a host is by examining the three-dimensional localization volume
associated with the GW event. As originally noted
by~\citet{1986Natur.323..310S} and later expanded
by~\citet{2012PhRvD..86d3011D}, this approach can be used in a statistical 
fashion: although any individual GW event may have many potential host galaxies
within the relevant localization volume, by analyzing many events simultaneously
statistical properties of the host galaxies (such as their redshifts) can be
inferred.
An essential aspect of this approach is the size of the localization
volume: the smaller the volume, the smaller the number of potential host
galaxies contained within, and the easier the statistical task of identifying
the true host. Previous work has indicated that these volumes are
{$>10^{4}\,\mbox{Mpc}^3$}, corresponding to {$>$hundreds} of 
potential galaxies~\citep{2013ApJ...767..124N,2014ApJ...784....8H,2016ApJ...820..136G,2016ApJ...829L..15S,2016arXiv160504242S}.

We re-examine the localization volumes associated with sources identified by
ground-based gravitational-wave detector networks. We perform a systematic
study, incorporating a range of networks and a range of potential sources, and
simulate tens of thousands of GW detections. 
This systematic study takes advantage of the rapid GW localization algorithm
described in~\citet{2015arXiv150900055C} and expanded to 3D in~\S\ref{sec:methods}. 
We focus on the loudest events, exploiting the fact that
the distribution of detected signal-to-noise ratios is
universal~\citep{2014arXiv1409.0522C} and that loud events can be anticipated
and are inevitable, especially these events are generally well-localized~\citep{2015arXiv150900055C}.
We find that the best localized events, which we call ``golden'' events,\footnote{Similar
but not identical to the events discussed in~\citet{2005ApJ...623..689H}, which
considered high SNR events for LISA as powerful tests of strong-field general relativity.}
can be constrained to very small volumes, potentially containing only a single
galaxy. 
For these golden events the unique host galaxy may be
identified, allowing a direct EM association with the GW source.

We describe our localization algorithm in~\S\ref{sec:methods}. In~\S\ref{sec:results} we provide an
overview of the results, focusing on the dependence of localization volume and area on
the GW detector network and the source properties. In~\S\ref{sec:discussion} we summarize our
results and conclude.
\section{Methods}\label{sec:methods}
\begin{figure}[t]
\centering
\includegraphics[width=1.1\columnwidth]{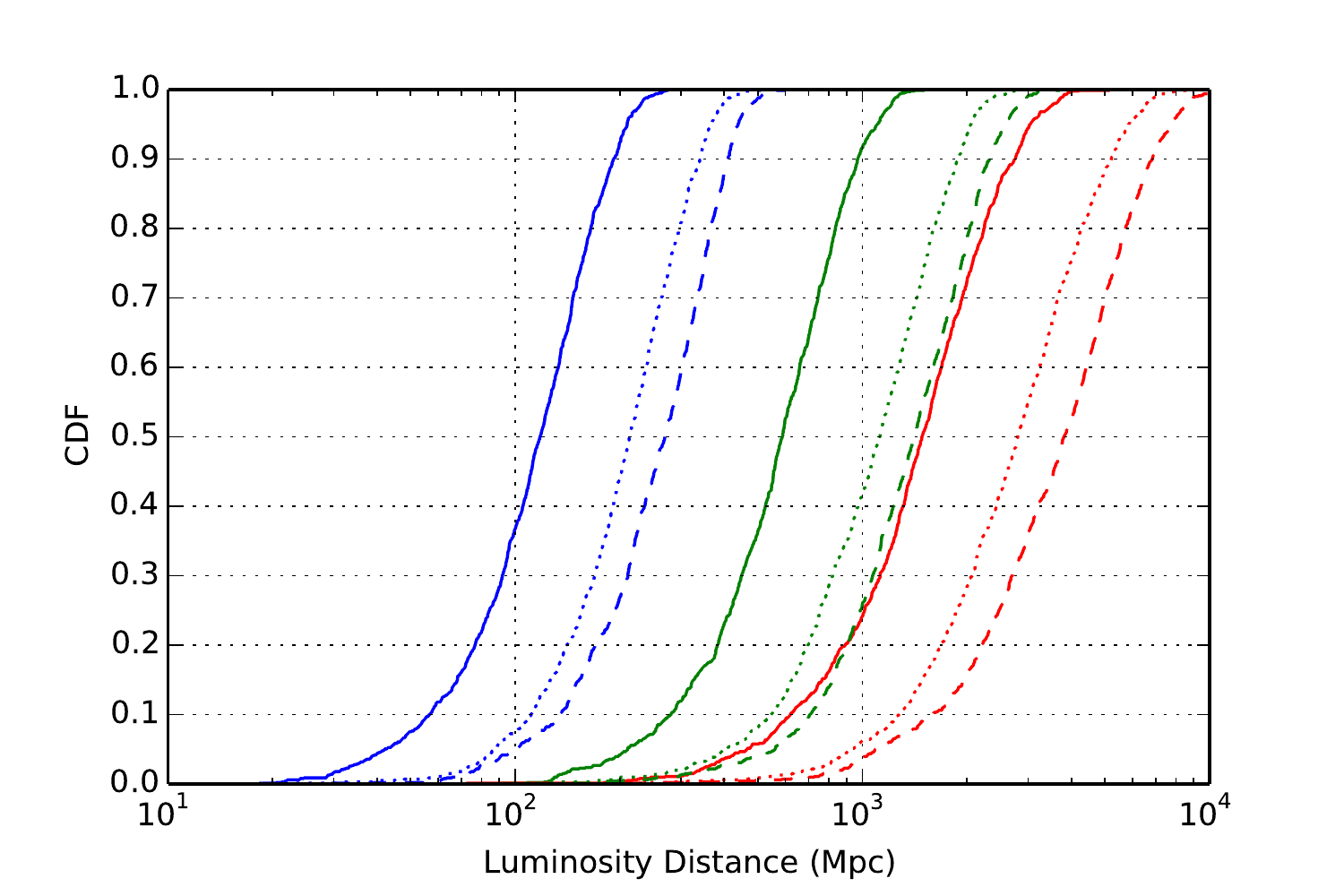}
\caption{\label{fig:culdist}
Cumulative distribution of the luminosity distances for the simulated detectable
1.4\msun--1.4\msun\ (blue), 10\msun--10\msun\ (green), and 30\msun--30\msun\ (red) binary mergers localized in this paper. 
We consider three GW detector network configurations at different
sensitivities: Solid--HLV
  at O3 sensitivity, dotted--HLV at design sensitivity, and dashed--HLVJI at
  design sensitivity. There are 10,000 simulated detectable/localized events for each type of binary systems and network configurations. 
}
\end{figure}

We take a Monte Carlo approach, generating a large sample of binary
  coalescences throughout the Universe and then investigating the fraction which
  can be detected by various GW networks, and producing the distributions of
  localizations associated with these detections.
We generate 1.4\msun--1.4\msun, 10\msun--10\msun, and 30\msun--30\msun\ binary
mergers at random sky positions with random inclinations and orientations.  The
waveforms of these mergers were generated using the waveform generator in
LALSuite (\url{https://wiki.ligo.org/DASWG/LALSuite}). For the 1.4\msun--1.4\msun\ binaries we use
the TaylorF2 waveform~\citep{2009PhRvD..80h4001Y}, and for the 10\msun--10\msun\ and 30\msun--30\msun\ binaries
we use the IMRPhenomD waveform~\citep{2016PhRvD..93d4007K}.  We assume aligned
spin and ignore precession; including general spins should not
  qualitatively change our results, and in general would lead to slightly
  improved localization~\citep{2016ApJ...825..116F}. We consider two GW detector network
configurations (HLV and HLVJI)\footnote{H: LIGO-Hanford, L: LIGO-Livingston, V:
  Virgo, J: KAGRA, and I: LIGO-India} at two different 
sensitivities (O3 and design).
The O3 sensitivity for H and L
corresponds to a binary neutron star (BNS) range of 120 Mpc, while for V the range is 60 Mpc~\citep{2016LRR....19....1A}. 
The {design} sensitivity for all detectors is
taken from the ``aLIGO'' curve in LIGO Document
T1500293 (\url{https://dcc.ligo.org/LIGO-T0900288/public}), corresponding to a
BNS range of 200~Mpc. Using these waveforms and detector sensitivities we calculate the
optimal match filter signal-to-noise ratio (SNR), $\rho$, at each detector:
\begin{equation}\label{eq:snr}
\rho=\left (\displaystyle\int_{f_{\rm low}}^{f_{\rm high}} \frac{|\tilde{h}(f)|^2}{S_h(f)} \, df \right )^{1/2},
\end{equation}
where $\tilde{h}(f)$ is the Fourier transform of the waveform in the frequency
domain and $S_h(f)$ is the detector power spectral density.  We add 
Gaussian noise of width 1 to the SNR, and calculate the network SNR as the
root-sum-square of the noise-added individual detector SNRs.  Our detection
threshold requires that the 
network SNR be greater than 12. 
The cumulative distribution of the luminosity distance of the detected events is
shown in Fig.~\ref{fig:culdist}; note the tails to nearby events for all scenarios. 
For detectable events
we calculate the time-of-arrival difference and phase difference between pairs
of detectors, and add in Gaussian errors following the Fisher matrix formalism~\citep{1994PhRvD..49.2658C}.

In order to localize the source we follow a similar approach to our previous work~\citep{2015arXiv150900055C}. We use the measured difference in arrival time,
$\Delta t$, the measured difference in phase, $\Delta \eta$, and the measured SNR in
the individual detectors to reconstruct the 3D location of the source, $(\theta,\phi,D_L)$:
\begin{equation}\label{eq:posterior}
f(\theta,\phi,D_L|\Delta t,\Delta \eta,\rho)=\frac{f(\Delta t,\Delta \eta,\rho|\theta,\phi,D_L)\, f(\theta,\phi,D_L)}{f(\Delta t,\Delta \eta,\rho)}.
\end{equation}
In our simulations the binary mergers are distributed uniformly in comoving
volume and the cosmological parameters are taken to be
$(\Omega_m=0.27,\Omega_\Lambda=0.73,h=0.71)$.  Our prior, $f(\theta,\phi,D_L)$,
assumes the same cosmological parameters, and is also uniform in comoving volume. 
To calculate the posterior in
Eq.~\ref{eq:posterior}, we grid the sky using healpix pixels (\url{http://healpix.jpl.nasa.gov}) 
and 400 bins in luminosity distance.  We scale the resolution in sky direction
and luminosity distance depending on the measured SNR of the events. Higher SNR
events have finer resolution (e.g., $N_{\rm side}=1024$ for network SNR of 110).
The likelihood function $f(\Delta t,\Delta \eta,\rho|\theta,\phi,D_L)$ is
estimated using similar $\chi^2$ likelihood methods to those presented
in~\citet{2015arXiv150900055C}.
For each binary above the network detection threshold we produce a joint likelihood in
sky position and distance. For each voxel (3D volume pixel) we produce a
probability. We then rank order the probabilities, and starting with the
highest probability voxel we sum until we have reached the desired probability
threshold. In this manner we produce the 90\% likelihood volumes presented in
\S\ref{sec:results}. We take the 3D likelihood and project it to obtain the
2D sky localization. Similarly, we then rank the 2D pixels to produce the 90\%
likelihood localization areas.

In order to verify our algorithm we have produced a ``P-P plot'', similar to the
verification presented in~\citet{2015arXiv150900055C}, and found our algorithm is working as expected. 
We have also compared our algorithm to LALInference~\citep{2015PhRvD..91d2003V}, a code designed to sample the posterior distribution 
for all source parameters using full models of the source waveform and all detector data, and find 
good agreement.
\begin{figure}[t]
\centering
\includegraphics[width=1.1\columnwidth]{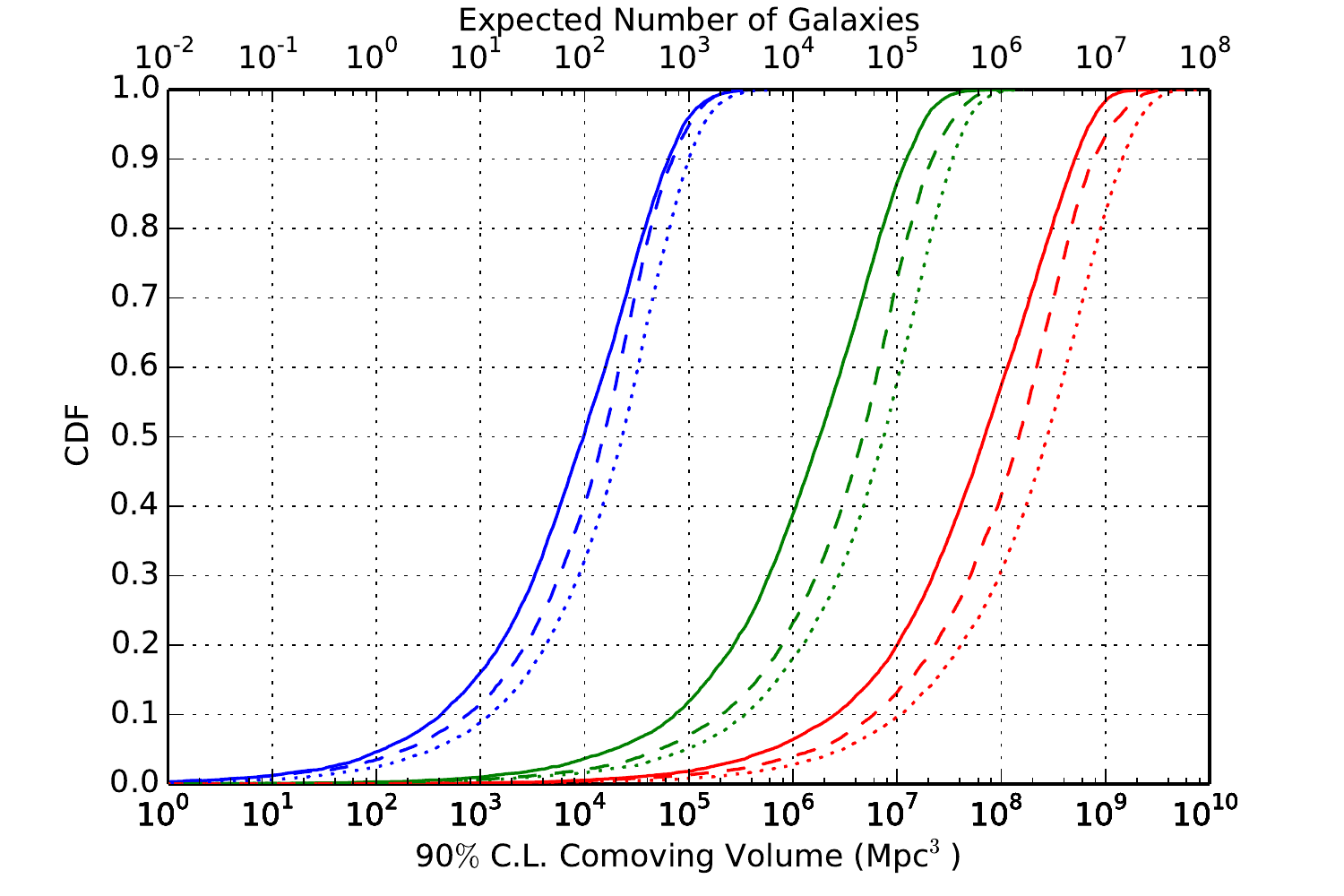}
\caption{\label{fig:culvol}
Cumulative distribution of the 90$\%$ confidence level 3D localization comoving
volumes for 10,000 simulated 
1.4\msun--1.4\msun\ (blue), 10\msun--10\msun\ (green), and 30\msun--30\msun\ (red) binary mergers. 
We consider three GW detector network configurations at different
sensitivities: Solid--HLV
  at O3 sensitivity, dotted--HLV at design sensitivity, and dashed--HLVJI at
  design sensitivity. The top axis shows the expected number of galaxies within the volume assuming the number density of galaxies 
is 0.01/Mpc$^3$
}
\end{figure} 
\section{results}
\label{sec:results}
Our results are summarized in Figs.~\ref{fig:culvol} and~\ref{fig:cularea} and Tables~\ref{table:volume} and~\ref{table:area}.
Fig.~\ref{fig:culvol} shows a plot of the cumulative distribution of
the 90\% localization volumes, while Fig.~\ref{fig:cularea} presents the cumulative distribution of
90\% localization sky areas. For example,
we find that 50\% of the BNS sources will be localized to a
volume of $\sim 10^4\,\mbox{Mpc}^3$, regardless of detector network.

We are particularly interested in the number of galaxies that can be expected in
these volumes. To estimate this, we assume that the number density of galaxies
is {0.01/Mpc$^3$}. This estimate takes the Schechter function~\citep{1976ApJ...203..297S} 
parameters in B-band $\phi_{*}=1.6\times 10^{-2}h^3$ Mpc$^{-3}$, $\alpha$=-1.07, $L_*=1.2\times 10^{10}h^{-2}L_{B,\odot}$
and $h=0.7$ (~\citet{2002MNRAS.336..907N,2003MNRAS.344..307L,2006A&A...445...51G,2016ApJ...820..136G}, $L_{B,\odot}$ is the solar luminosity in B-band), 
integrating down to $0.12\,L^*$ 
and comprising 86\% of the total luminosity.   
The top axis in Fig.~\ref{fig:culvol} shows the expected
number of galaxies in the 90\% localization volumes. 
For example, a localization
volume of 1000~Mpc$^3$ corresponds to an expectation of 10 galaxies, while a localization
smaller than 100 Mpc$^3$ corresponds to on average a single galaxy within the localization
volume. In other words, for GW sources localized to within 100 $\mbox{Mpc}^3$, it may
be possible to directly identify the host galaxy without the need for an
associated EM transient.

We have argued that a small fraction of GW sources are sufficiently well
localized to allow for improved  constraints on their host galaxies.
But how many
of these systems will actually be detected? To estimate this we need to know the 
expected number of systems that will be detected by the various
networks. With only a few systems detected to date, the rate of binary coalescences remains uncertain; we assume the rate for 
(1.4, 1.4),  (10, 10), and (30, 30) \msun\ binary mergers to be [low, mean, high]: 
$[10^{-8},10^{-6},10^{-5}]$, $[1\times 10^{-8},5\times 10^{-8},2\times 10^{-7}]$, and $[6\times 10^{-9},2\times 10^{-8}, 6\times 10^{-8}]$ Mpc$^{-3}$yr$^{-1}$, 
respectively~\citep{2010CQGra..27q3001A,2016arXiv160203842T,2016PhRvX...6d1015A,2016arXiv160707456T}.
From these rates we are able to estimate the number of systems
that will be detected with each GW network, and then use the results in Fig.~\ref{fig:culvol} to infer the number of
systems that will be localized sufficiently to identify the unique host
galaxy. We summarize our results in Tables~\ref{table:volume} and~\ref{table:area}.
\begin{table*}[tp]%
\centering
\begin{tabular}{cclccccc}
\toprule
\hline\hline
       Network &Mass(\msun)     & Median (Mpc$^3$) 	& $<100$ Mpc$^3$ ($\%$)	& $<100$ Mpc$^3$ ($\left<{N_{\rm event}}\right>$)	& $<1000$ Mpc$^3$ ($\%$)	& $<1000$ Mpc$^3$ ($\left<{N_{\rm event}}\right>$)\\ \midrule
\hline
        HLV O3 &(1.4, 1.4) 			& $9.8\times 10^3$    	& 4.6	&[0.0024, 0.24, 2.4]	&16.1		&[0.0083, 0.83, 8.3]						\\ 
        HLV O3 &(10, 10) 			& $1.8\times 10^6$    	& 0.3	&[0.017, 0.085, 0.34]	&1.0		&[0.061, 0.34, 1.2]					\\ 
        HLV O3 &(30, 30) 			& $6.9\times 10^7$    	& 0.1	&[0.021, 0.070, 0.21]	&0.1		&[0.055, 0.18, 0.55]					\\ 
						
       	HLV design &(1.4, 1.4)  			& $2.3\times 10^4$    	& 2.4	&[0.0078, 0.78, 7.8]	&8.9		&[0.029, 2.9, 29]					\\ 
       	HLV design &(10, 10)  			& $7.4\times 10^6$    	& 0.2	&[0.077, 0.38, 1.5]	&0.6		&[0.21, 1.0, 4.2]					\\ 
       	HLV design &(30, 30)  			& $2.8\times 10^8$    	& 0.0	&[0.067, 0.22, 0.67]	&0.1		&[0.27, 0.89, 2.7]					\\ 

        HLVJI design &(1.4, 1.4)			& $1.5\times 10^4$    	& 3.5	&[0.023, 2.3, 23]	&11.5		&[0.076, 7.6, 76]					\\	
        HLVJI design &(10, 10)			& $4.6\times 10^6$    	& 0.2	&[0.13, 0.66, 2.6]	&0.7		&[0.51, 2.5, 10]					\\	
        HLVJI design &(30, 30)			& $1.5\times 10^8$    	& 0.0	&[0.043, 0.14, 0.43]	&0.1		&[0.60, 2.0, 6.0]					\\
\hline\hline
\end{tabular}
\caption{\label{table:volume}
90$\%$ confidence level 3D localization comoving volumes. First column: detector network and sensitivity. 
Second column: mass of the binaries. Third column: median 90$\%$ confidence level localization volume. 
Fourth \& sixth columns: fraction of events localized within 100 and 1000~Mpc$^3$. 
Fifth \& seventh columns: expected number of events localized within 100 and 1000~Mpc$^3$ for [low, mean, high] event rate densities, where the intrinsic rate densities of (1.4, 1.4),  (10, 10), and (30, 30) \msun\ binary mergers are taken to be 
$[10^{-8},10^{-6},10^{-5}]$, $[1\times 10^{-8},5\times 10^{-8},2\times 10^{-7}]$, and $[6\times 10^{-9},2\times 10^{-8}, 6\times 10^{-8}]$ Mpc$^{-3}$yr$^{-1}$, respectively. 
The run duration for O3 and design are taken to be 9 months and 1 year. The duty
cycle is not considered in this calculation (i.e.,it is taken to be
100\%). 
}
\end{table*}
There are a number of interesting aspects of our results:
\begin{itemize}
\begin{figure}[b]
    \centering
    \includegraphics[width=1.1\columnwidth]{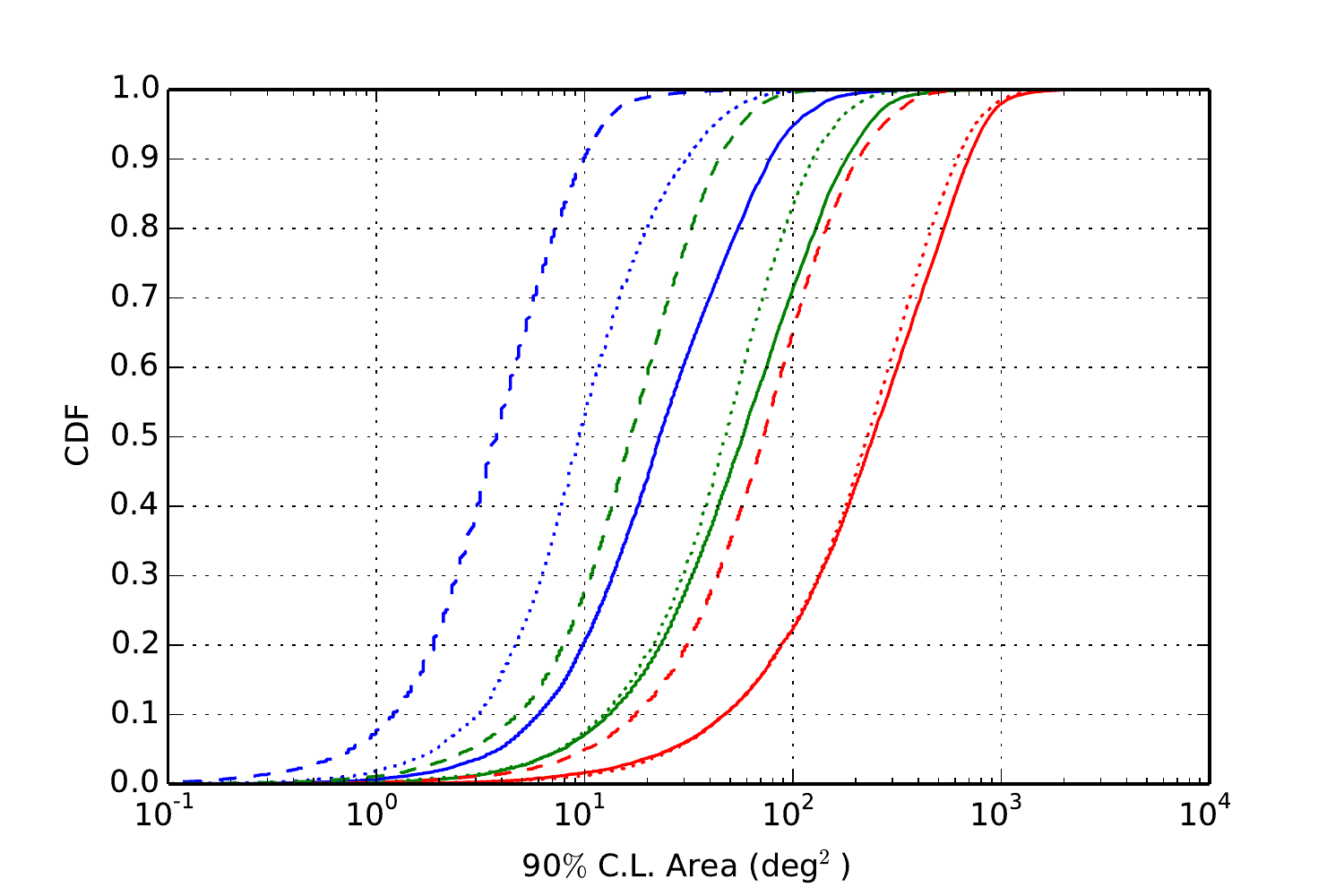}  
\caption{\label{fig:cularea}
Cumulative distribution of the 90$\%$ confidence level 2D localization areas for
10,000 simulated 
1.4\msun--1.4\msun\ (blue), 10\msun--10\msun\ (green), and 30\msun--30\msun\ (red) binary mergers. 
We consider three GW detector network configurations at different
sensitivities: Solid--HLV
  at O3 sensitivity, dotted--HLV at design sensitivity, and dashed--HLVJI at
  design sensitivity. 
}
\end{figure}
\item The localization volumes are larger for more massive binaries. This is
  because these binaries are generally detected to higher distances, and
  therefore a fractional error in distance or sky position corresponds to a
  larger volume. Massive binaries also merge at lower frequency and spend shorter time in the
  GW detectable band, leading to a larger error in estimates of the time-of-arrival.
  We find that BNS systems
  are the most likely to be well localized, and therefore are the most likely to
  allow for unique host galaxy association.
\item  The localization areas are larger for more massive binaries. 
   Massive binaries are detected at higher distances, so their signals are significantly redshifted and, 
   on top of their intrinsic lower merging frequency, spend even shorter 
   time in band. This is why, for example, the 10-10\msun\ binary black hole (BBH) localization area is $\sim$2.5 times that of the BNSs at HLV O3 sensitivity, 
   with the ratio increasing to $\sim$5 at design sensitivity.
\item The localization areas are smaller for improved gravitational-wave
  networks. This is because the localization is primarily dependent on the
timing measurements at each detector, and as the networks improve the relative
timing improves. Going from three detectors to five detectors causes a large
improvement in the 2D localization.
\item In all cases there is a tail to very well localized events, corresponding
  to events that are particularly loud and are fortuitously oriented and located
  on the sky. For example, for BNS events localized within 1000 $\,\mbox{Mpc}^3$ by HLV design sensitivity, 
  the average measured network SNR is 37.
\item We find that $\sim16\%$ of BNS sources detected with the HLV network in O3
  will be localized to within 1,000$\,\mbox{Mpc}^3$, corresponding to $<10$
  galaxies within the 90\% localization volume. One in 20 sources will be
  localized to within 100$\, \mbox{Mpc}^3$, indicating that a unique host galaxy
  may be identified for these events. These values become 2\%/9\% (at 100/1,000$\,\mbox{Mpc}^3$) for HLV at
  design sensitivity, and 4\%/12\% at HLVJI at design sensitivity.
\item We find that a large fraction of BNS systems are 2D localized to better than
  10 deg$^2$ for all network configurations; for HLVJI over 90\% of sources are
  localized this well, with almost 10\% localized to within 1
  deg$^2$. The population of well-localized 2D systems overlaps
  significantly with the population of well-localized 3D systems; these are the
  high-SNR tail of the universal distribution.
\item As the networks improve the fraction of well-localized events can drop as
  increasing numbers of sources are detected at increasing distance. However,
  the total number of golden events will increase, since a golden event for a
  given network will only become ``more golden'' when detected by an improved network.
\item We find that for all network configurations there is an expectation of
  $\sim1$ golden event per year: a binary localized to a sufficiently small volume that a
  unique galaxy host may be identified. 
\end{itemize}

\section{Discussion}
\label{sec:discussion}
\begin{table*}[]%
\centering
\begin{tabular}{cclccccc}
\toprule
\hline\hline
       Network &Mass(\msun)     & Median (deg$^2$) 	& $<1$ deg$^2$ ($\%$)	& $<1$ deg$^2$ ($\left<{N_{\rm event}}\right>$)	& $<10$ deg$^2$ ($\%$)	& $<10$ deg$^2$ ($\left<{N_{\rm event}}\right>$)\\ \midrule
\hline
        HLV O3 &(1.4, 1.4) 			& 23    	& 0.6	&[0.00033, 0.033, 0.33]	&20.4			&[0.011, 1.1, 11]					\\ 
        HLV O3 &(10, 10) 			& 57    	& 0.2	&[0.012, 0.058, 0.23]	&7.1			&[0.43, 2.1, 8.6]					\\ 
        HLV O3 &(30, 30) 			& 242    	& 0.1	&[0.021, 0.070, 0.21]	&1.6			&[0.68, 2.3, 6.8]					\\ 
						
       	HLV design &(1.4, 1.4)  			& 9    	& 2.0	&[0.0065, 0.65, 6.5]	&52.5			&[0.17, 17, 170]				\\ 
       	HLV design &(10, 10)  			& 47    	& 0.3	&[0.12, 0.59, 2.3]	&7.5		&[2.8, 14, 55]					\\ 
       	HLV design &(30, 30)  			& 228    	& 0.0	&[0.089, 0.30, 0.89]	&1.2		&[2.6, 8.7, 26]					\\ 

        HLVJI design &(1.4, 1.4)			& 4    	& 7.8	&[0.052, 5.2, 52]	&90.2			&[0.60, 60, 597]				\\	
        HLVJI design &(10, 10)			& 17    	& 1.1	&[0.83, 4.1, 17]	&27.5		&[20, 101, 403]					\\	
        HLVJI design &(30, 30)			& 72    	& 0.2	&[0.99, 3.3, 9.9]	&4.9		&[21, 71, 212]					\\
\hline\hline
\end{tabular}
\caption{\label{table:area}
90$\%$ confidence level 2D localization areas. First column: detector network and sensitivity. 
Second column: mass of the binaries. Third column: median 90$\%$ confidence level localization area. 
Fourth \& sixth columns: fraction of events localized within 1 and 10~deg$^2$. 
Fifth \& seventh columns: expected number of events localized within 1 and 10~deg$^2$ for [low, mean, high] event rate densities. 
}
\end{table*}
We have explored the ability of gravitational-wave detector networks to localize
sources both on the sky and in space (sky localization + distance). We have
derived the distribution of expected localizations for a
range of detector networks and a range of sources. Of particular interest are
the cases where the expected number of galaxies is
$\left<{N}\right>=1$, since in these cases it is theoretically
possible to identify the {\em unique}\/ host galaxy associated with a GW
sources. We show that in all network configurations there exists a small percentage
of sources that will be sufficiently localized so that only a single galaxy may
be present in the localization volume. We find that even in O3 there is the likelihood of
an event sufficiently well localized such that only a few galaxies will be found
within its localization volume, and in the advanced detector era
we expect tens of sources per year localized to only a very small
number of potential host galaxies, and to be able to uniquely identify the host galaxy of a few
sources per year. 
Figs.~\ref{fig:culvol} and~\ref{fig:cularea} show that localizations deteriorate quickly beyond the
first few golden events, suggesting that the utility of this small sample of
events may outweigh the rest of the population.

There are a number of potential complications which need to be discussed. We
have ignored the role of systematic errors in the GW measurements. For example,
a 5\% systematic error in the distance estimates will lead to a 5\% shift in the
90\% confidence volume, and therefore potentially the wrong associated host
galaxies. We have performed preliminary tests to show that, given the large
spread in distance uncertainties, these effects remain sub-dominant at
the 5\% level. We have ignored gravitational lensing, since the best localized sources
are generally the closest ones. For example, even for the 30--30 \msun\ case, the
maximum luminosity distance of our sample of binaries localized to within 1,000 Mpc$^3$ is
$\sim$400 Mpc. This corresponds to a redshift of $\sim$0.1; gravitational lensing is
expected to be negligible at this distance.
We note that GW detectors localize sources in luminosity
distance, while galaxies generically live in redshift space. In the work above
we have assumed a given cosmology to perform this mapping. Since one
of our goals is to use these systems as standard sirens to measure
cosmological parameters, one may be concerned that our cosmological assumptions 
preclude independent cosmological measurements. However, if we 
conservatively assume that the cosmological parameters are known to within 10\%,
it may nonetheless be possible to narrow down 
the number of host galaxies. These can then be used in turn to further constrain
the cosmological parameters to better than 10\% through the use of standard
sirens. We have tested this explicitly. 
In addition, we have ignored the effects of galaxy
clustering. Certainly some volumes may center on clusters with a galaxy density orders
of magnitude higher than the average. In these cases, identifying the individual
host galaxy may become more challenging, although estimating the redshift
of the cluster may nonetheless be possible.

Another important complication is the absence of complete galaxy catalogs within
the localization volumes of interest. However, we have shown that the associated
2D sky localizations are small for events with well-localized volumes. 
For example,
for BNS events localized by the HLV network at design sensitivity with $\left<{N}\right>\sim10$ galaxies within
their localization volume, the median 90\% sky localization area is 1.8 deg$^2$.
We
emphasize that large-field survey telescopes such as the Dark Energy Camera (DECam), Hyper Suprime-Cam, and LSST would be
able to fully cover these fields in a small number of pointings. Furthermore, as
mentioned above, these well localized systems are the closest ones in the
population, and therefore are expected to have correspondingly brighter
host galaxies. 
For example, for the maximum distance of the 30--30 \msun\ case discussed above (400
Mpc), the B-band magnitude of a Milky Way-like galaxy would be $\sim17.6$. 
Thus an instrument such as DECam would be able to build a complete galaxy catalog
{\em in real time}\/ across the entire localization region in a few short
($\sim1\,\mbox{minute}$) pointings.
In addition to building an on-the-fly galaxy catalog, the small number of
potential host galaxies within the volume coupled with the comparatively nearby
distances for these well-localized sources will enable unparalleled triggered
searches for EM counterparts. This is further enabled because our 3D
localization algorithm presented in \S\ref{sec:methods}, as well as the BAYESTAR
algorithm presented in~\citet{2016ApJ...829L..15S}, can be run in low
latency ($\sim\,\mbox{minutes}$).

Of particular interest is how these better-localized binaries will
impact attempts to measure the Hubble constant with gravitational-wave standard
sirens. As far as standard sirens are concerned, identifying a host galaxy is
equivalent to identifying an EM counterpart: both cases allow for an independent
measurement of the redshift. Although only a small fraction of the sources will
be well-localized, since these directly allow for points on the Hubble diagram
they provide uniquely powerful constraints~\citep{2006PhRvD..74f3006D,2010ApJ...725..496N}.
We note that the known localization does not significantly improve the distance
estimates~\citep{2016arXiv161005633P}, although we are in the process of exploring this conclusions
explicitly for our sample of golden events.

In conclusion, we have shown that there will exist a class of events
which will be sufficiently well localized to allow us to narrow the potential
number of host galaxies to a small number, and in extreme cases allow for the
unique identification of the host of the GW source. These special cases will
directly engender multi-messenger astronomy without the need for associated
transient EM counterparts. These golden events are likely to be of particular
interest in the development of GW astrophysics and cosmology.

\begin{acknowledgments}
We acknowledge valuable discussions with Salvatore Vitale and Benjamin Farr. 
We also acknowledge Will Farr for very helpful comments.
The authors were supported by NSF CAREER grant
PHY-1151836. They were also supported in part by the Kavli Institute for
Cosmological Physics at the University of Chicago through NSF grant PHY-1125897
and an endowment from the Kavli Foundation. 
The authors acknowledge the University of Chicago Research Computing Center for support of this work.
\end{acknowledgments}

\bibliography{references}

\begin{thebibliography}{}
\expandafter\ifx\csname natexlab\endcsname\relax\def\natexlab#1{#1}\fi

\bibitem[{{Abadie} {et~al.}(2010){Abadie}, {Abbott}, {Abbott}, {Abernathy},
  {Accadia}, {Acernese}, {Adams}, {Adhikari}, {Ajith}, {Allen}, \&
  et~al.}]{2010CQGra..27q3001A}
{Abadie}, J., {Abbott}, B.~P., {Abbott}, R., {et~al.} 2010, Classical and
  Quantum Gravity, 27, 173001

\bibitem[{{Abbott} {et~al.}(2016{\natexlab{a}}){Abbott}, {Abbott}, {Abbott},
  {Abernathy}, {Acernese}, {et~al.}}]{2016arXiv160707456T}
{Abbott}, B.~P., {Abbott}, R., {Abbott}, T.~D., {et~al.} 2016{\natexlab{a}},
  arXiv:1607.07456, arXiv:1607.07456

\bibitem[{{Abbott} {et~al.}(2016{\natexlab{b}}){Abbott}, {Abbott}, {Abbott},
  {Abernathy}, {Acernese}, {Ackley}, {Adams}, {Adams}, {Addesso}, {Adhikari},
  \& et~al.}]{2016ApJ...818L..22A}
---. 2016{\natexlab{b}}, Astrophys. J. Lett., 818, L22

\bibitem[{{Abbott} {et~al.}(2016{\natexlab{c}}){Abbott}, {Abbott}, {Abbott},
  {Abernathy}, {Acernese}, {Ackley}, {Adams}, {Adams}, {Addesso}, {Adhikari},
  \& et~al.}]{2016PhRvX...6d1015A}
---. 2016{\natexlab{c}}, Physical Review X, 6, 041015

\bibitem[{{Abbott} {et~al.}(2016{\natexlab{d}}){Abbott}, {Abbott}, {Abbott},
  {Abernathy}, {Acernese}, {Ackley}, {Adams}, {Adams}, {Addesso}, {Adhikari},
  \& et~al.}]{2016PhRvL.116x1103A}
---. 2016{\natexlab{d}}, Physical Review Letters, 116, 241103

\bibitem[{{Abbott} {et~al.}(2016{\natexlab{e}}){Abbott}, {Abbott}, {Abbott},
  {Abernathy}, {Acernese}, {Ackley}, {Adams}, {Adams}, {Addesso}, {Adhikari},
  \& et~al.}]{2016ApJ...826L..13A}
---. 2016{\natexlab{e}}, Astrophys. J. Lett., 826, L13

\bibitem[{{Abbott} {et~al.}(2016{\natexlab{f}}){Abbott}, {Abbott}, {Abbott},
  {Abernathy}, {Acernese}, {Ackley}, {Adams}, {Adams}, {Addesso}, {Adhikari},
  \& et~al.}]{2016PhRvL.116f1102A}
---. 2016{\natexlab{f}}, Physical Review Letters, 116, 061102

\bibitem[{{Abbott} {et~al.}(2016{\natexlab{g}}){Abbott}, {Abbott}, {Abbott},
  {Abernathy}, {Acernese}, {Ackley}, {Adams}, {Adams}, {Addesso}, {Adhikari},
  \& et~al.}]{2016LRR....19....1A}
---. 2016{\natexlab{g}}, Living Reviews in Relativity, 19, arXiv:1304.0670

\bibitem[{{Abbott} {et~al.}(2016{\natexlab{h}}){Abbott}, {Abbott}, {Abbott},
  {Abernathy}, {Acernese}, {Ackley}, {Adams}, {Adams}, \&
  et~al.}]{2016arXiv160203842T}
---. 2016{\natexlab{h}}, arXiv:1602.03842, arXiv:1602.03842

\bibitem[{{Annis} {et~al.}(2016){Annis}, {Soares-Santos}, {Berger}, {Brout},
  {Chen}, {Chornock}, {Cowperthwaite}, {Diehl}, {Doctor}, {Drlica-Wagner},
  {Drout}, {Farr}, {Finley}, {Flaugher}, {Foley}, {Frieman}, {Gruendl},
  {Herner}, {Holz}, {Kessler}, {Lin}, {Marriner}, {Neilsen}, {Rest}, {Sako},
  {Smith}, {Smith}, {Sobreira}, {Walker}, {Yanny}, {Abbott}, {Abdalla},
  {Allam}, {Benoit-L{\'e}vy}, {Bernstein}, {Bertin}, {Buckley-Geer}, {Burke},
  {Capozzi}, {Carnero Rosell}, {Carrasco Kind}, {Carretero}, {Castander},
  {Cenko}, {Crocce}, {Cunha}, {D'Andrea}, {da Costa}, {Desai}, {Dietrich},
  {Eifler}, {Evrard}, {Fernandez}, {Fischer}, {Fong}, {Fosalba}, {Fox},
  {Fryer}, {Garcia-Bellido}, {Gaztanaga}, {Gerdes}, {Goldstein}, {Gruen},
  {Gutierrez}, {Honscheid}, {James}, {Karliner}, {Kasen}, {Kent}, {Kuehn},
  {Kuropatkin}, {Lahav}, {Li}, {Lima}, {Maia}, {Martini}, {Metzger}, {Miller},
  {Miquel}, {Mohr}, {Nichol}, {Nord}, {Ogando}, {Peoples}, {Petravic},
  {Plazas}, {Quataert}, {Romer}, {Roodman}, {Rykoff}, {Sanchez}, {Santiago},
  {Scarpine}, {Schindler}, {Schubnell}, {Sevilla-Noarbe}, {Sheldon}, {Smith},
  {Stebbins}, {Swanson}, {Tarle}, {Thaler}, {Thomas}, {Tucker}, {Vikram},
  {Wechsler}, {Weller}, {Wester}, \& {DES Collaboration}}]{2016ApJ...823L..34A}
{Annis}, J., {Soares-Santos}, M., {Berger}, E., {et~al.} 2016, Astrophys. J.
  Lett., 823, L34

\bibitem[{{Chen} \& {Holz}(2014)}]{2014arXiv1409.0522C}
{Chen}, H.-Y., \& {Holz}, D.~E. 2014, arXiv:1409.0522, arXiv:1409.0522

\bibitem[{{Chen} \& {Holz}(2015)}]{2015arXiv150900055C}
---. 2015, arXiv:1509.00055, arXiv:1509.00055

\bibitem[{{Cowperthwaite} {et~al.}(2016){Cowperthwaite}, {Berger},
  {Soares-Santos}, {Annis}, {Brout}, {Brown}, {Buckley-Geer}, {Cenko}, {Chen},
  {Chornock}, {Diehl}, {Doctor}, {Drlica-Wagner}, {Drout}, {Farr}, {Finley},
  {Foley}, {Fong}, {Fox}, {Frieman}, {Garcia-Bellido}, {Gill}, {Gruendl},
  {Herner}, {Holz}, {Kasen}, {Kessler}, {Lin}, {Margutti}, {Marriner},
  {Matheson}, {Metzger}, {Neilsen}, {Quataert}, {Rest}, {Sako}, {Scolnic},
  {Smith}, {Sobreira}, {Strampelli}, {Villar}, {Walker}, {Wester}, {Williams},
  {Yanny}, {Abbott}, {Abdalla}, {Allam}, {Armstrong}, {Bechtol},
  {Benoit-L{\'e}vy}, {Bertin}, {Brooks}, {Burke}, {Carnero Rosell}, {Carrasco
  Kind}, {Carretero}, {Castander}, {Cunha}, {D'Andrea}, {da Costa}, {Desai},
  {Dietrich}, {Evrard}, {Fausti Neto}, {Fosalba}, {Gerdes}, {Giannantonio},
  {Goldstein}, {Gruen}, {Gutierrez}, {Honscheid}, {James}, {Johnson},
  {Johnson}, {Krause}, {Kuehn}, {Kuropatkin}, {Lima}, {Maia}, {Marshall},
  {Menanteau}, {Miquel}, {Mohr}, {Nichol}, {Nord}, {Ogando}, {Plazas}, {Reil},
  {Romer}, {Sanchez}, {Scarpine}, {Sevilla-Noarbe}, {Smith}, {Suchyta},
  {Tarle}, {Thomas}, {Thomas}, {Tucker}, {Weller}, \& {DES
  Collaboration}}]{2016ApJ...826L..29C}
{Cowperthwaite}, P.~S., {Berger}, E., {Soares-Santos}, M., {et~al.} 2016,
  Astrophys. J. Lett., 826, L29

\bibitem[{{Cutler} \& {Flanagan}(1994)}]{1994PhRvD..49.2658C}
{Cutler}, C., \& {Flanagan}, {\'E}.~E. 1994, \prd, 49, 2658

\bibitem[{{Dalal} {et~al.}(2006){Dalal}, {Holz}, {Hughes}, \&
  {Jain}}]{2006PhRvD..74f3006D}
{Dalal}, N., {Holz}, D.~E., {Hughes}, S.~A., \& {Jain}, B. 2006, \prd, 74,
  063006

\bibitem[{{Del Pozzo}(2012)}]{2012PhRvD..86d3011D}
{Del Pozzo}, W. 2012, \prd, 86, 043011

\bibitem[{{Farr} {et~al.}(2016){Farr}, {Berry}, {Farr}, {Haster}, {Middleton},
  {Cannon}, {Graff}, {Hanna}, {Mandel}, {Pankow}, {Price}, {Sidery}, {Singer},
  {Urban}, {Vecchio}, {Veitch}, \& {Vitale}}]{2016ApJ...825..116F}
{Farr}, B., {Berry}, C.~P.~L., {Farr}, W.~M., {et~al.} 2016, \apj, 825, 116

\bibitem[{{Gehrels} {et~al.}(2016){Gehrels}, {Cannizzo}, {Kanner}, {Kasliwal},
  {Nissanke}, \& {Singer}}]{2016ApJ...820..136G}
{Gehrels}, N., {Cannizzo}, J.~K., {Kanner}, J., {et~al.} 2016, \apj, 820, 136

\bibitem[{{Gonz{\'a}lez} {et~al.}(2006){Gonz{\'a}lez}, {Lares}, {Lambas}, \&
  {Valotto}}]{2006A&A...445...51G}
{Gonz{\'a}lez}, R.~E., {Lares}, M., {Lambas}, D.~G., \& {Valotto}, C. 2006,
  \aap, 445, 51

\bibitem[{{Hanna} {et~al.}(2014){Hanna}, {Mandel}, \&
  {Vousden}}]{2014ApJ...784....8H}
{Hanna}, C., {Mandel}, I., \& {Vousden}, W. 2014, \apj, 784, 8

\bibitem[{{Hughes} \& {Menou}(2005)}]{2005ApJ...623..689H}
{Hughes}, S.~A., \& {Menou}, K. 2005, \apj, 623, 689

\bibitem[{{Khan} {et~al.}(2016){Khan}, {Husa}, {Hannam}, {Ohme}, {P{\"u}rrer},
  {Forteza}, \& {Boh{\'e}}}]{2016PhRvD..93d4007K}
{Khan}, S., {Husa}, S., {Hannam}, M., {et~al.} 2016, \prd, 93, 044007

\bibitem[{{Liske} {et~al.}(2003){Liske}, {Lemon}, {Driver}, {Cross}, \&
  {Couch}}]{2003MNRAS.344..307L}
{Liske}, J., {Lemon}, D.~J., {Driver}, S.~P., {Cross}, N.~J.~G., \& {Couch},
  W.~J. 2003, \mnras, 344, 307

\bibitem[{{Nissanke} {et~al.}(2010){Nissanke}, {Holz}, {Hughes}, {Dalal}, \&
  {Sievers}}]{2010ApJ...725..496N}
{Nissanke}, S., {Holz}, D.~E., {Hughes}, S.~A., {Dalal}, N., \& {Sievers},
  J.~L. 2010, Astrophys. J., 725, 496

\bibitem[{{Nissanke} {et~al.}(2013){Nissanke}, {Kasliwal}, \&
  {Georgieva}}]{2013ApJ...767..124N}
{Nissanke}, S., {Kasliwal}, M., \& {Georgieva}, A. 2013, \apj, 767, 124

\bibitem[{{Norberg} {et~al.}(2002){Norberg}, {Cole}, {Baugh}, {Frenk},
  {Baldry}, {Bland-Hawthorn}, {Bridges}, {Cannon}, {Colless}, {Collins},
  {Couch}, {Cross}, {Dalton}, {De Propris}, {Driver}, {Efstathiou}, {Ellis},
  {Glazebrook}, {Jackson}, {Lahav}, {Lewis}, {Lumsden}, {Maddox}, {Madgwick},
  {Peacock}, {Peterson}, {Sutherland}, {Taylor}, \& {2DFGRS
  Team}}]{2002MNRAS.336..907N}
{Norberg}, P., {Cole}, S., {Baugh}, C.~M., {et~al.} 2002, \mnras, 336, 907

\bibitem[{{Pankow} {et~al.}(2016){Pankow}, {Sampson}, {Perri}, {Chase},
  {Coughlin}, {Zevin}, \& {Kalogera}}]{2016arXiv161005633P}
{Pankow}, C., {Sampson}, L., {Perri}, L., {et~al.} 2016, ArXiv e-prints,
  arXiv:1610.05633

\bibitem[{{Schechter}(1976)}]{1976ApJ...203..297S}
{Schechter}, P. 1976, \apj, 203, 297

\bibitem[{{Schutz}(1986)}]{1986Natur.323..310S}
{Schutz}, B.~F. 1986, \nat, 323, 310

\bibitem[{{Singer} {et~al.}(2016{\natexlab{a}}){Singer}, {Chen}, {Holz},
  {Farr}, {Price}, {Raymond}, {Cenko}, {Gehrels}, {Cannizzo}, {Kasliwal},
  {Nissanke}, {Coughlin}, {Farr}, {Urban}, {Vitale}, {Veitch}, {Graff},
  {Berry}, {Mohapatra}, \& {Mandel}}]{2016ApJ...829L..15S}
{Singer}, L.~P., {Chen}, H.-Y., {Holz}, D.~E., {et~al.} 2016{\natexlab{a}},
  \apjl, 829, L15

\bibitem[{{Singer} {et~al.}(2016{\natexlab{b}}){Singer}, {Chen}, {Holz},
  {Farr}, {Price}, {Raymond}, {Cenko}, {Gehrels}, {Cannizzo}, {Kasliwal},
  {Nissanke}, {Coughlin}, {Farr}, {Urban}, {Vitale}, {Veitch}, {Graff},
  {Berry}, {Mohapatra}, \& {Mandel}}]{2016arXiv160504242S}
---. 2016{\natexlab{b}}, Astrophys. J.s, 226, 10

\bibitem[{{Soares-Santos} {et~al.}(2016){Soares-Santos}, {Kessler}, {Berger},
  {Annis}, {Brout}, {Buckley-Geer}, {Chen}, {Cowperthwaite}, {Diehl}, {Doctor},
  {Drlica-Wagner}, {Farr}, {Finley}, {Flaugher}, {Foley}, {Frieman}, {Gruendl},
  {Herner}, {Holz}, {Lin}, {Marriner}, {Neilsen}, {Rest}, {Sako}, {Scolnic},
  {Sobreira}, {Walker}, {Wester}, {Yanny}, {Abbott}, {Abdalla}, {Allam},
  {Armstrong}, {Banerji}, {Benoit-L{\'e}vy}, {Bernstein}, {Bertin}, {Brown},
  {Burke}, {Capozzi}, {Carnero Rosell}, {Carrasco Kind}, {Carretero},
  {Castander}, {Cenko}, {Chornock}, {Crocce}, {D'Andrea}, {da Costa}, {Desai},
  {Dietrich}, {Drout}, {Eifler}, {Estrada}, {Evrard}, {Fairhurst}, {Fernandez},
  {Fischer}, {Fong}, {Fosalba}, {Fox}, {Fryer}, {Garcia-Bellido}, {Gaztanaga},
  {Gerdes}, {Goldstein}, {Gruen}, {Gutierrez}, {Honscheid}, {James},
  {Karliner}, {Kasen}, {Kent}, {Kuropatkin}, {Kuehn}, {Lahav}, {Li}, {Lima},
  {Maia}, {Margutti}, {Martini}, {Matheson}, {McMahon}, {Metzger}, {Miller},
  {Miquel}, {Mohr}, {Nichol}, {Nord}, {Ogando}, {Peoples}, {Plazas},
  {Quataert}, {Romer}, {Roodman}, {Rykoff}, {Sanchez}, {Scarpine}, {Schindler},
  {Schubnell}, {Sevilla-Noarbe}, {Sheldon}, {Smith}, {Smith}, {Smith},
  {Stebbins}, {Sutton}, {Swanson}, {Tarle}, {Thaler}, {Thomas}, {Tucker},
  {Vikram}, {Wechsler}, {Weller}, \& {DES Collaboration}}]{2016ApJ...823L..33S}
{Soares-Santos}, M., {Kessler}, R., {Berger}, E., {et~al.} 2016, Astrophys. J.
  Lett., 823, L33

\bibitem[{{Veitch} {et~al.}(2015){Veitch}, {Raymond}, {Farr}, {Farr}, {Graff},
  {Vitale}, {Aylott}, {Blackburn}, {Christensen}, {Coughlin}, {Del Pozzo},
  {Feroz}, {Gair}, {Haster}, {Kalogera}, {Littenberg}, {Mandel},
  {O'Shaughnessy}, {Pitkin}, {Rodriguez}, {R{\"o}ver}, {Sidery}, {Smith}, {Van
  Der Sluys}, {Vecchio}, {Vousden}, \& {Wade}}]{2015PhRvD..91d2003V}
{Veitch}, J., {Raymond}, V., {Farr}, B., {et~al.} 2015, \prd, 91, 042003

\bibitem[{{Yunes} {et~al.}(2009){Yunes}, {Arun}, {Berti}, \&
  {Will}}]{2009PhRvD..80h4001Y}
{Yunes}, N., {Arun}, K.~G., {Berti}, E., \& {Will}, C.~M. 2009, \prd, 80,
  084001

\end{thebibliography}

\end{document}